\documentclass[prl,aps,twocolumn,twoside,english,superscriptaddress]{revtex4}

\usepackage{epsfig,verbatim}
\usepackage{graphicx}
\usepackage{amsmath}
\usepackage{revsymb}

\usepackage{amssymb}

\usepackage[caption=false]{subfig}

\def\dg{{\dagger}}

\newcommand{\ket}[1]{|  #1 \rangle}
\newcommand{\bra}[1]{ \langle #1  |}
\newcommand{\proj}[1]{\ket{#1}\!\bra{#1}}

\newcommand{\Tr}{{\mathrm{Tr}}}
\newcommand{\cN}{{\cal N}}
\newcommand{\cD}{{\cal D}}
\newcommand{\cE}{{\cal E}}
\newcommand{\cM}{{\cal M}}

\newcommand\bea{\begin{eqnarray}}
\newcommand\eea{\end{eqnarray}}
\newcommand{\beq}{\begin{equation}}
\newcommand{\eeq}{\end{equation}}

\newcommand{\Id}{\mathbb{I}}

\renewcommand{\section}[1]{\emph{#1}---}

\begin{document}

\title{\Large Detecting Incapacity}
\author{Graeme Smith}\email{gsbsmith@gmail.com}
\affiliation{IBM T.J. Watson Research Center, Yorktown Heights, NY
10598, USA}
\author{John A. Smolin}\email{smolin@watson.ibm.com}
\affiliation{IBM T.J. Watson Research Center, Yorktown Heights, NY
10598, USA}

\date{\today}

\begin{abstract}

  Using unreliable or noisy components for reliable communication
  requires error correction.  But which noise processes can support
  information transmission, and which are too destructive?  For
  classical systems any channel whose output depends on its input has
  the capacity for communication, but the situation is substantially
  more complicated in the quantum setting.  We find a generic test for
  incapacity based on any suitable forbidden transformation---a
  protocol for communication with a channel passing our test would
  also allow us to implement the associated forbidden transformation.
  Our approach includes both known quantum incapacity tests---positive
  partial transposition (PPT) and antidegradability (no cloning)---as
  special cases, putting them both on the same footing. We also find a
  physical principle explaining the nondistillability of PPT states:
  Any protocol for distilling entanglement from such a state would
  also give a protocol for implementing the forbidden time-reversal
  operation.

 \end{abstract}

\maketitle

Understanding how to use communication resources optimally is a major
part of information theory \cite{Shannon48,CoverThomas}.  Usually, a
sender and receiver have access to repeated uses of a noisy
communication channel and want to use error correction to introduce
redundancy into their messages, making them less susceptible to noise.
The capacity of a channel, measured in bits per channel use, is the
best rate that can be achieved with vanishing probability of
transmission error in the limit of many channel uses.

Quantum information theory expands the notions of both information and noise to include quantum effects \cite{BS04}.
Indeed, a quantum channel has many different capacities, depending on what kind of information is to be transmitted, and
what other resources are available.  For example, the classical capacity of a quantum channel is the best rate for transmitting
classical information, the private capacity relates to quantum cryptography, and the quantum capacity is
 the best rate for coherent communication of quantum systems.  This last, which establishes the fundamental limits on quantum error 
 correcting codes, will be our main concern here.

\begin{figure}[htbp]
\includegraphics[width=3in]{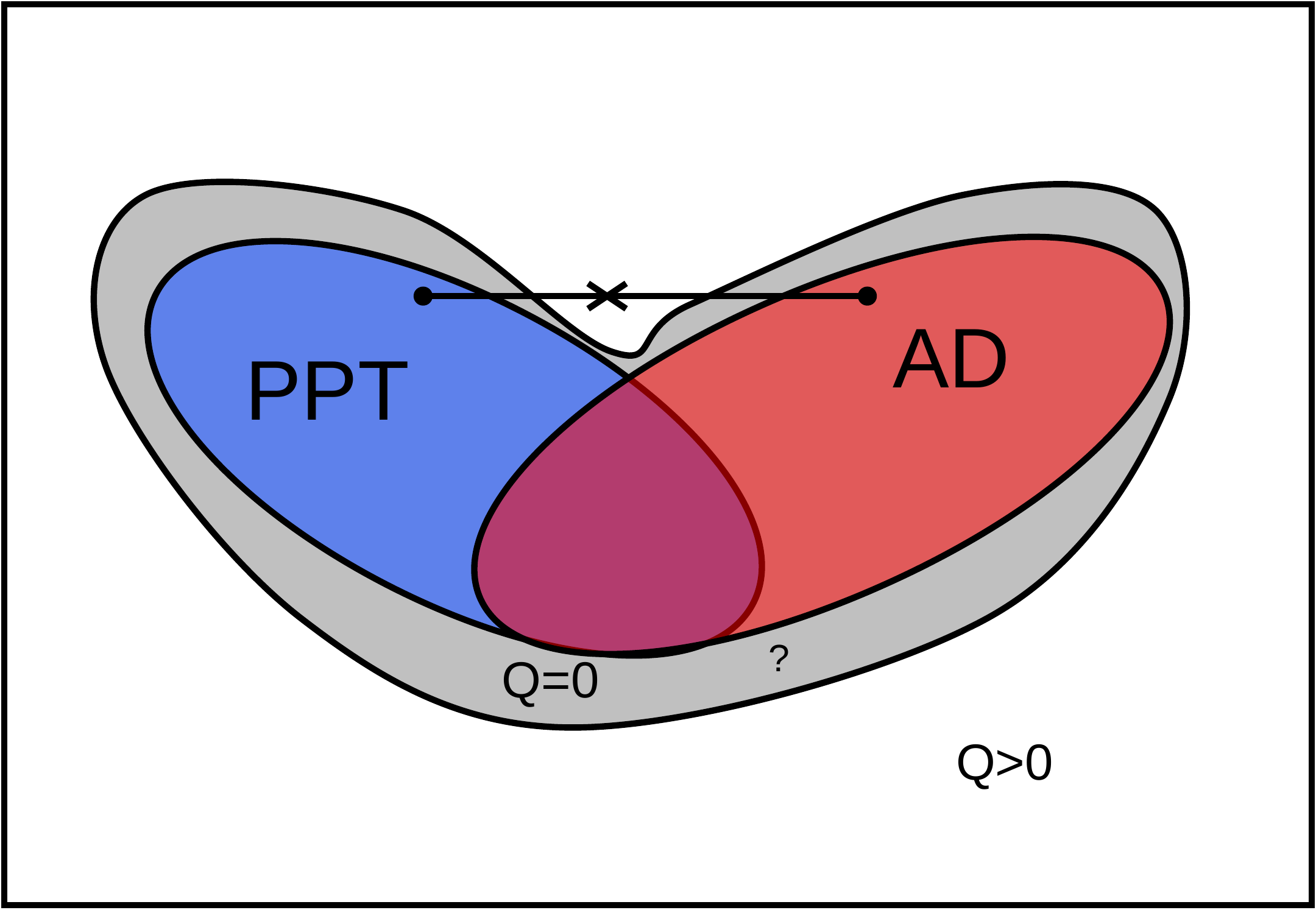}
\caption{Zero quantum capacity channels.  There are two known
  incapacity tests, and two associated known sets of zero quantum
  capacity channels.  Antidegradable channels have no capacity because
  otherwise they could be used to violate no cloning.  Positive
  partial transpose channels satisfy the mathematical criterion that
  their Choi matrices remain positive under partial transposition.  It
  is possible to find pairs of zero quantum capacity channels, one PPT
  and the other Antidegradable, that have positive joint quantum
  capacity.  This can be used to show that the set of zero quantum
  capacity channels is non convex.  It is unknown whether there are
  zero quantum capacity channels that are neither antidegradable nor
  PPT.  In this paper we find that the reason PPT channels have no
  quantum capacity is that any capacity would allow them to implement
  the forbidden time reversal operation.  }
\end{figure}

In contrast to the simple capacity formula of classical information
theory, finding the quantum capacity of a quantum channel is in
general intractable.  The formula for the quantum capacity \cite{Lloyd97,Shor02,D03}
of a channel involves a maximization over states on tensor products of an
arbitrary number of uses of the channel.  No bound on the size of this maximization
is yet known.  Even the \emph{classical} capacity of a quantum channel has this
problem \cite{Hastings09}.

We will consider the simpler incapacity question: which channels have
nonzero quantum capacity, and which don't?  This is still very hard
to answer, but there are two known criteria for showing a channel has
no capacity.  The first, a mathematical condition called the positive
partial transpose (PPT) criterion, is due to \cite{HHH96,Peres96}.
The second, due to \cite{Erase97,BDEFMS98}, gives a condition for when
quantum capacity would imply a violation of the no-cloning principle
\cite{WZ82}.  Adding to the complexity of the zero-quantum-capacity
channels is the existence of the superactivation phenomenon, where two
zero-capacity channels interact synergistically to jointly generate
capacity \cite{SY08}.  A particular consequence of this phenomenon is
that the set of zero-capacity quantum channels is not convex.
Unfortunately, there is no simple test to
determine, given its parameters, whether a channel can transmit quantum information.

We will address the incapacity question from an abstract point of
view.  Our arguments require only the existence of physical states and
maps, as well as the existence of a suitable forbidden (or unphysical) map on
the states. One motivation of this work is to understand incapacity of quantum
channels, but our approach is sufficiently general to include such
things as generalized probabilistic theories \cite{Barrett07} as
well as the discrete quantum mechanics of \cite{Modal2010}.  Our
findings will also apply to classical systems with proscribed operations. 

\section{Preliminaries}
\begin{figure}[htbp]
\includegraphics[width=3in]{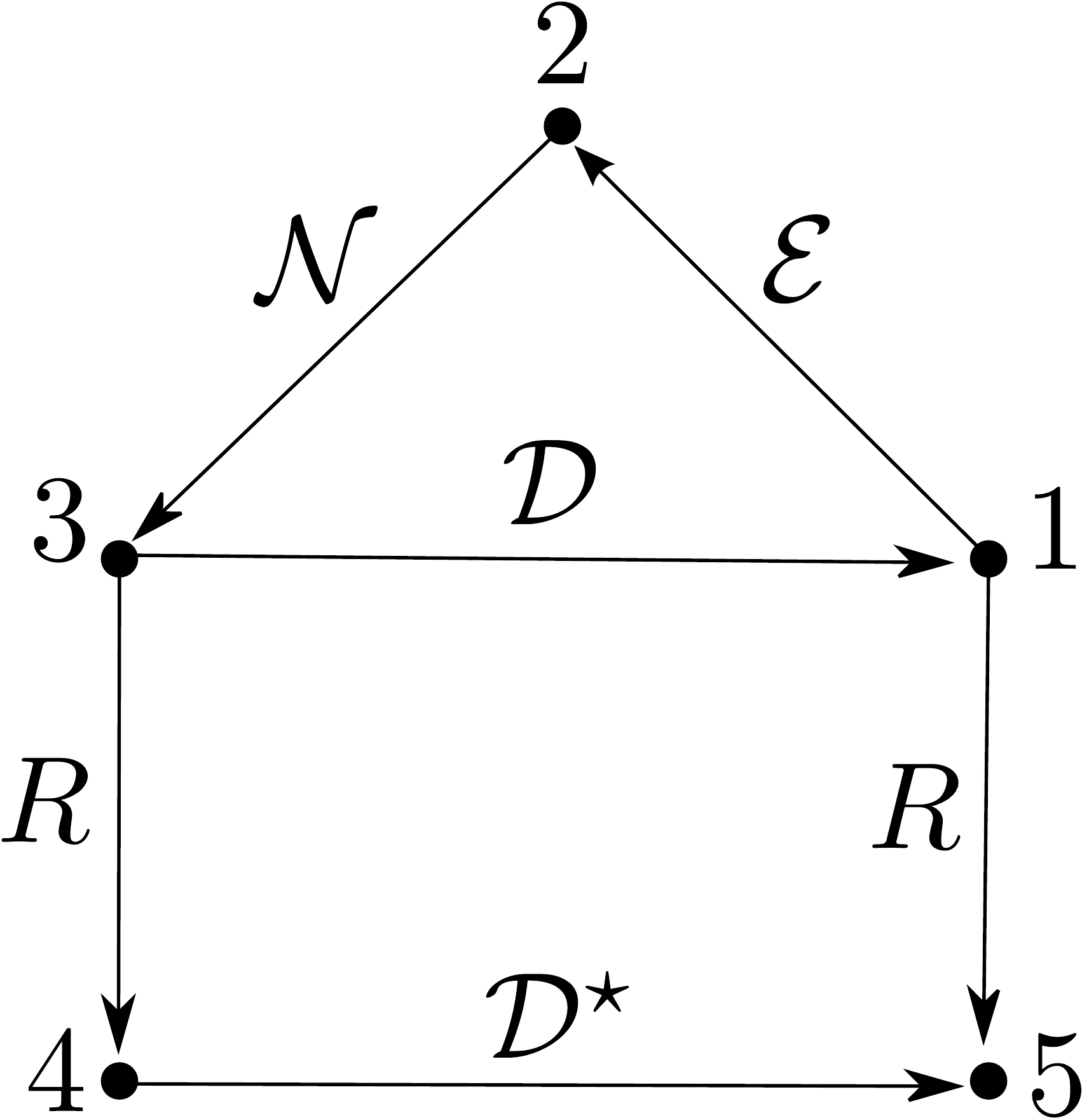} \\

\caption{Sketch of Lemma 1:   The assumptions that go into the lemma  
are illustrated by the connections on this graph.  ${\mathbf P}$-commutation
is represented by $3\rightarrow 1\rightarrow 5$ leading to the same result
as $3 \rightarrow 4 \rightarrow 5$.  The existence of encoder and decoder
that transmit states successfully though $\cN$ is indicated by the
path $1 \rightarrow 2 \rightarrow 3\rightarrow 1$ which is the identity.
We also assume that the path $R \circ \cN$ or $2 \rightarrow 3\rightarrow 4$ 
is physical even though $R$ is not.  Then the nonphysical operation 
$R$ ($1\rightarrow 5$) is implemented by the physical path 
$1\rightarrow 2 \rightarrow 3 \rightarrow 4 \rightarrow 5$.
}
\label{Fig:Involution}
\end{figure}

The theories we will consider have minimal structure.  We assume a
physical state space $B$ and a set of allowed physical operations $\mathbf{P}$
from $B\rightarrow B$ that is closed under composition.  For quantum
mechanics, $B$ will be the set of density matrices and $\mathbf{P}$ will
be the set of trace-preserving completely positive maps.  We will also require
a nonphysical operation $R: B\rightarrow B$ with $R\notin \mathbf{P}$.
This $R$ will need the following crucial property:

\textbf{Definition 1}:  ($\mathbf{P}$-commutation) An unphysical map $R$
is $\mathbf{P}$-commutative if for every 
$\cD \in \mathbf{P}$ there is a $\cD^\star \in \mathbf{P}$ such that
$R\circ \cD = \cD^{\star} \circ R$.  See Fig. \ref{Fig:Involution}.

Note that for any unphysical $R$ there must be a set of states $S\subset B$
such that no $\cN \in {\mathbf P}$ has for all $\phi \in S$ $R(\phi) =
\cN(\phi)$.  We say that $R$ is unphysical on $S$.  The following
simple lemma will be remarkably useful (see Figure~2).

\textbf{Lemma 1}: If $R$ is unphysical on $S$ and is
$\mathbf{P}$-commutative, then any $\cN$ with $R \circ \cN \in
{\mathbf P}$ cannot reliably transmit the states in $S$.

\textbf{Proof:} Suppose there were physical encoding and decoding
operations that allowed transmission of states in $S$.  In other
words, there are physical $\cE$ and $\cD$ such that
\begin{equation}
\psi = (\cD \circ \cN \circ \cE)(\psi)\ \ \forall \psi \in S.  
\end{equation}
Then
\begin{equation}
R(\psi) = R\circ\cD \circ \cN \circ \cE(\psi) = \cD^{\star}\circ R \circ \cN \circ \cE (\psi)
\label{commutedit}
\end{equation}
which, recalling that $\cM = R \circ \cN$ is physical, gives a
prescription for physically implementing $R$ on $S$ as $\cD^{\star}
\circ \cM \circ \cE$, which is impossible by hypothesis. $\Box$

Note that this argument implies that if $R$ is continuous, and there
is no physical operation that approximates $R$ to high precision on
$S$ then $\cN$ cannot transmit the states in $S$ to high precision.

\section{Application to quantum mechanics}
The rest of the paper is dedicated to extending and exploring the
consequences of this line of reasoning, focusing on its application to
quantum mechanics.  While a quantum channel $\cN$ implicitly acts only
on some particular part of $B$ of dimension $d$, our unphysical maps
will be defined for any input dimension.  So, the unphysical map can
be thought of as a family of maps acting on different input
dimensions.  To streamline notation we will suppress labels indicating
this dimension whenever possible.

Our usual notion of capacity is defined over many uses of a channel,
${\cal N}^{\otimes n}$, and we will also need to consider nonphysical
maps acting on $d^n$, which we call $R^{(n)}$.  The parentheses
indicate that this map may not be a tensor product of maps $R$ on the
individual systems, but can act jointly on all $n$ systems.
For linear maps, the tensor product is well defined and in this case
we will typically take $R^{(n)}=R^{\otimes n}$.

We are now ready to consider a simple example of Lemma 1 applied to
the time reversal, or transpose, operation $T$.  Recall that time
reversal is an operation that preserves inner products between states,
but is anti-unitary rather than unitary.  It is also the canonical
example of a positive map that is not completely positive
\footnote{Recall that a positive map $\gamma$ has the property that
  $\gamma(\rho) \ge 0$, for all $\rho \ge 0$ but not necessarily $(\Id
  \otimes \gamma)(\rho') \ge 0$ for some $\rho' \ge 0$ on a larger
  space.  A completely positive map $\eta$ has $(\Id \otimes
  \eta)(\rho') \ge 0$ for all $\rho'\ge 0$ \protect\cite{choi}}.  As
such, the transpose maps states to states, but is not a physically
implementable operation.  In fact, letting $S$ be the set of all qubit
states, time-reversal is unphysical on $S$.  Now, given a channel
$\cD(\rho) = \sum_i A_i \rho A_i^\dagger$ letting $\cD^{\star}(\rho) =
\sum_i A_i^* \rho (A_i^{*})^\dagger$ (with $*$ denoting complex
conjugation), it is easy to check that $\cD^{\star}$ is also a channel
and furthermore $T\circ \cD = \cD^{\star} \circ T$.  Having
demonstrated $\mathbf{P}$-commutativity, we can now apply Lemma 1.  In
fact,
it is easy to show that if $T \circ \cN$ is a quantum channel, so is
$T \circ \cN^{\otimes n}$ for any $n$, so we see from Lemma 1 that any
channel with $T\circ \cN$ physical must have zero capacity.

Translating this to a statement about the channel's Choi matrix \footnote{
The Choi matrix of a channel $\cN$ completely characterizes the channel and
is defined as $(I \otimes T)((I\otimes \cN)(\proj{\phi_d}))$, where
$\ket{\phi_d}=\frac{1}{\sqrt{d}}\sum_{i=1}^d \ket{i}\ket{i}$.}, we
find that any channel $\cN$ such that $(I \otimes T)((I\otimes
\cN)(\proj{\phi_d})) \geq 0$ (in other words a PPT Choi matrix) must
have zero quantum capacity.  The \emph{reason} is that any capacity for
such a channel would lead to a protocol for implementing the
(unphysical) time-reversal operation.  A similar argument shows that
any protocol for distilling pure entanglement from a PPT state could
also be used to implement time-reversal (this is shown in the
appendix).

There being no reason to restrict to the transpose operation, we
can easily see that given any continuous  $\mathbf{P}$-commuting $R^{(n)}$ which 
cannot be approximated by a physical map on qubit states, then any $\cN$ with   
$R^{(n)} \circ \cN^{\otimes n} \in {\mathbf P}$ 
has zero quantum capacity.

\section{Positive linear maps}
If we take $R$ to be a linear map that preserves system dimension, it
is possible to completely characterize the zero-capacity channels
detected by our method.  This is achieved through the following
theorem originally communicated to us by Choi
\cite{choiprivatecommunication}.  We present here an alternative proof
that makes an explicit connection to the theory of group
representations.

\textbf{Theorem 1:} Take states and channels to be the usual quantum
mechanics with $d$-level systems.  If $R$ is linear, invertible,
preserves system dimension and trace, and is $\mathbf{P}$-commutative
it is either of the form $R(\rho) = (1-p)\rho^T + pI/d$ or $R(\rho) =
(1-p)\rho + pI/d$.

\textbf{Proof:} Let $R$ be such a map, and consider the requirements of
${\mathbf P}$-commutation.  Defining $\cN_U(\rho) = U \rho U^\dagger$, we have
$\cN^{\star}_U = R \circ \cN_U \circ R^{-1}$ and see immediately that
since $\cN_U$ is invertible, so is $\cN^{\star}_U$.  Since it
preserves dimension and is a physical channel, $\cN^{\star}_U$ must be
simply conjugation by a unitary $V_U$.  Furthermore,
$\cN^{\star}_{U_1U_2} = R \circ \cN^{\star}_{U_1U_2} \circ R^{-1} = R
\circ \cN^{\star}_{U_1} \circ R^{-1} \circ R \circ \cN^{\star}_{U_2}
\circ R^{-1} = \cN^{\star}_{U_1} \circ \cN^{\star}_{U_2}$, so that
$V_U$ must be a d-dimensional representation of $\mathbb{U}(d)$.
Since $\cN_U = R^{-1} \circ \cN_U^{\star} \circ R$, this
representation must be faithful (\emph{i.e.}, invertible) up to an
overall phase which will cancel under conjugation. There are only two
such representations: the fundamental and complex conjugate
representations.

Suppose we have $\cN^{\star}_U = \cN_U$.  Then, for all $\rho$, $U^\dg
R(U\rho U^\dg)U = R(\rho)$, so that letting $\ket{\phi_d} =
\frac{1}{\sqrt{d}}\sum_{i=1}^d \ket{i}\ket{i}$ and
$\proj{\phi_d}^{\Gamma}$ denote its partial transpose, we have 
\begin{equation}
\nonumber U^\dg
\otimes U^\dg (I \otimes R)(\proj{\phi_d}^{\Gamma}) U \otimes U = (I
\otimes R)(\proj{\phi_d}^{\Gamma}).  
\end{equation}
As a result, we have 
\begin{equation}
\nonumber (I \otimes R)(\proj{\phi_d}^{\Gamma}) = a I +b F\ \mathrm{where}\ 
F=\sum_{i,j} \ket{i}\!\ket{j}\!\bra{j}\!\bra{i}
\end{equation} 
so that $(I \otimes R)(\proj{\phi_d}) =
aI + db\proj{\phi_d}$.  Requiring $R$ to preserve trace, we thus find
$R(\rho) = p I/d + (1-p)\rho$.

We now consider $\cN^{\star}_U = \cN_{U^*}$, which works similarly.
Specifically, we now have $(I \otimes R)(\proj{\phi_d}) = a I + b F$
so that, requiring $R$ to preserve trace, we find $R(\rho) = pI/d +
(1-p)\rho^T.$ $\Box$

\section{Decoder-dependent $R$ and no cloning}
The above theorem shows that the only nonphysical $\mathbf{P}$-commutative
linear maps
are of the form $R(\rho)=(1-p)\rho^T+pI/d$.  In
order to make statements about capacities, $R^{\otimes n}$ must also
be  $\mathbf{P}$-commutative.  The only such nonphysical map is $R(\rho)
=\rho^T$. As a result detecting incapacity in non-PPT channels
requires $R$ to be nonlinear.

In order to accommodate nonlinear maps, we must generalize our ideas
about $\mathbf{P}$-commutation.  Specifically, we introduce the following
definition of $\mathbf{P}$-commutation for a class of unphysical operations.

\textbf{Definition 2}: ($\mathbf{P}$-commutation) A set of unphysical maps
$\{R_{\cD}\}_{\cD \in \mathbf{P}}$ with $R_{\cD} \notin \mathbf{P}$ is 
$\mathbf{P}$-commutative if for all 
maps $\cD \in \mathbf{P}$ there is a $\cD^\star \in \mathbf{P}$ and an
unphysical $R$ with $R_{\cD}\circ \cD = \cD^{\star} \circ R$.


In Eq.~(\ref{commutedit}), we were able to commute $R$ and $\cD$ by
replacing $\cD$ with $\cD^\star$.  The motivation behind our new 
definition is to allow more freedom in finding decoders that remain
physical after commutation with $R$, at the cost of having
channel-dependent nonphysical maps $R_\cD$.  Note that if $R$ is
invertible, $\cD^\star= R_\cD \circ\cD\circ R^{-1}$ will define
$\star$.

Now, given a $\mathbf{P}$-commuting family of maps $\{R_{\cD}\}$ that are unphysical 
on a qubit with
associated $\star$ and $R$, any channel $\cN$ for which $R \circ \cN$
is physical cannot transmit all states in some set $S$ reliably.
Within quantum mechanics, given $R^{(n)}_{\cD}$, $R^{(n)}$, and
$\star$ then if $R^{(n)} \circ \cN^{\otimes n}$ is physical, $\cN$ has
no quantum capacity.

In quantum mechanics, any decoder can be written in terms of a unitary
$U$ as $\cD(\psi)=\Tr_E U (\psi\otimes\proj{0}_E) U^\dag \approx \psi$
which implies $U (\psi\otimes\proj{0}_E) U^\dag \approx
\psi\otimes\sigma$ for some $\sigma$ independent of $\psi$.  We
therefore only need consider the simpler set of nonphysical maps $R_U$
with $U^\star R(\psi\otimes\proj{0}_E) (U^\star)^\dag =R_U (U
\psi\otimes\proj{0}_E U^\dag)$ defining $\star$ to detect the
incapacity of any $\cN$ with $R \circ (\cN \otimes \proj{0}_E)$
physical.

Every quantum channel has an isometric
extension to an environment.  An antidegradable channel is a channel
for which the environment, by further processing, can simulate the
original channel \cite{DS03,GF04}.  As a result of the no cloning
theorem, such channels can be shown have zero quantum capacity
\cite{Erase97,BDEFMS98}.  We now use the nonphysical cloning
operation to give a simple proof that antidegradable
channels have zero quantum capacity.

If $\cM$ is antidegradable with input $A$ and output $B$, then there
is an extension of $\cM$, $\cM_{12}$ from $A$ to $B_1B_2$, such that
for all $\rho$, $\Tr_{B_2}\cM_{12}(\rho) = \Tr_{B_1}\cM_{12}(\rho) =
\cM(\rho)$.  Any $\psi$ on $B$ has a unique decomposition $\psi =
\tilde{\psi} + \sigma$, with $\tilde{\psi}$ in the range of $\cM$ and
$\sigma$ in its orthogonal complement.  Now define $\tilde{R}(\psi) =
\cM_{12}\circ \cM^{-1}(\tilde{\psi}) + \sigma \otimes \sigma$ where
the pseudo-inverse $\cM^{-1}$ maps $\tilde{\psi}$ to its unique
preimage in the orthogonal complement of the kernel of $\cM$.  This
$\tilde{R}$ is continuous and $\Tr_1 \tilde{R}(\psi) =
\Tr_2\tilde{R}(\psi) = \psi$.  Furthermore, $\tilde{R}\circ \cM =
\cM_{12}$ is physical.  We can similarly extend $\tilde{R}$ to a continuous
$R$ from $BE$ to $B_1B_2E_1E_2$ with 
$R\circ (\cM \otimes \proj{0}_E) = \cM_{12} \otimes \proj{0}_{E_1}\otimes \proj{0}_{E_2}$
and $\Tr_1 {R}(\psi) =\Tr_2{R}(\psi) = \psi$.

Now, let $U^\star = U \otimes U$.
Choosing
$R_U(\rho_{BE})=U\otimes U R(U^\dag \rho_{BE} U) (U^\dag\otimes
U^\dag)$ and $\cN_U(\rho) = U \rho U^\dg$, we have $R_U\circ \cN_U =
\cN_{U^\star}\circ R$ and $R\circ (\cM \otimes \proj{0}_E )= \cM_{12}
\otimes \proj{0}\otimes\proj{0}$.  $\Tr_1 R_U(\psi \otimes
\proj{0})=\Tr_2 R_U(\psi \otimes \proj{0}) = \psi \otimes \proj{0}$,
so $R_{U}$ can clone an arbitrary state and is therefore unphysical.

Having demonstrated a nonphysical $R_U$, a $\star$ and physical $R
\circ (\cM \otimes \proj{0}_E)$, we have so far shown that a single use of
an anti-degradable channel can't be used to transmit a quantum state
with high fidelity.  However, since the tensor product of two
anti-degradable channels is again anti-degradable, this also shows
that many copies cannot transmit quantum information either.  As a
result, the capacity must be zero.

\section{Discussion}
We have presented a general approach for detecting the incapacity for
quantum communication using unphysical transformations, and shown that
both known incapacity tests fall into this framework. Furthermore we
have discovered a connection to the theory of representations of the
unitary group, and both positive partial transposition and
antidegradability correspond to simple representations.  This paves
the way for the discovery of new incapacity tests, and Theorem 1
suggests a fruitful direction, namely forbidden transformations that
are not linear.  


We have focused primarily on the standard (one-way) quantum capacity,
but these ideas can also be extended to the two-way capacity and
questions of entanglement distillation.  For example, in the appendix
we demonstrate non-distillability of PPT states
by showing that any successful distillation protocol could be used to
implement the unphysical time-reversal operation.  Our argument there
makes crucial use of the linearity of time reversal, which in light of
Theorem 1 severely restricts the detection of two-way capacity and
distillability.  Finding an argument relating two-way capacities and
nonlinear forbidden transformations is an important challenge.

We end on a speculative note.  We have shown that both known
incapacity tests are derived from fundamentally unphysical
transformations on state space: time reversal and cloning.  Could it
be that \emph{any} zero quantum capacity channel has such a ``reason'' for
its incapacity?  Formally, of course, the answer is ``yes''---the
forbidden transformation could just be a successful encoding or
decoding operation for the channel.  However, we would be much more
satisfied with something less tautological.  A good starting
point might be to identify a minimal set of primitive forbidden
operations that includes cloning and time-reversal as examples.

\emph{Acknowledgments:} We are grateful to Man-Duen Choi for telling
us about Theorem 1 and for very informative discussions.  Thanks also
to Toby Cubit and Robert Koenig for helpful suggestions and Charlie Bennett for advice 
on the manuscript.  This work
was supported by the DARPA QUEST program under contract
no. HR0011-09-C-0047.  


\vspace{0.05 in}
\section{Appendix:
Two-way capacities and distillable entanglement}  
We now argue that if the Choi matrix of a channel is PPT, it cannot be
distillable \cite{BDSW96} via local operation and classical communication (LOCC).
Since the Choi matrix of $\cN$ is PPT iff $T \circ \cN$ is a physical
channel, showing this will also prove that such a channel has no
$Q_2$, or quantum capacity assisted by LOCC.

By teleporting through the Choi matrix of a channel $\cN$, an LOCC
protocol with Kraus operators $A_i \otimes B_i$ can be used to prepare
the state
\begin{equation}\label{Eq:LOCC}
  \frac{d_{\rm out}}{d_{\rm in}} \sum_i \frac{1}{d_{\rm in}^2}\sum_u B_i \left( \cN (A_i^T \sigma_u \psi \sigma_u^\dg A_i^*)\right)B_i^\dg \otimes \proj{u}.
\end{equation}
where the $\sigma_u$s are generalized Pauli matrices.

Furthermore, if the Choi matrix can be distilled to a maximally
entangled state, there is an LOCC protocol such that
\begin{equation}
\psi = \frac{2}{d}\sum_i B_i \cN(A_i^T \psi A_i^*)B_i^\dg
\end{equation}
for an arbitrary qubit state $\psi$.  Letting $T(\rho) = \rho^T$, this implies that 
\begin{equation}
T(\psi) = \frac{2}{d}\sum_i B_i^* T\circ\cN(A_i^T \psi A_i^*)B_i^T,
\end{equation}
and 
\begin{equation}
T(\sigma_u\psi \sigma_u^\dg) = \frac{2}{d}\sum_i B_i^* T\circ\cN(A_i^T \sigma_u\psi \sigma_u^\dg A_i^*)B_i^T.
\end{equation}
This latter can be used to show that
\begin{equation}
T(\psi ) = \frac{2}{d}\sum_i \sigma_u^* B_i^* T\circ\cN(A_i^T \sigma_u\psi \sigma_u^\dg A_i^*)B_i^T\sigma_u^T
\end{equation}
the right hand of which, using the fact that $T\circ \cN$ is physical combined with Eq.~(\ref{Eq:LOCC}), gives 
a recipe for physically implementing $T$ using the LOCC operation with Kraus operators $A_i \otimes B_i^*$.
  As a result, the Choi matrix of $\cN$ must not be distillable.

\end{document}